# Decentralized Intelligence Network (DIN)

Abraham Nash
University of Oxford
abraham.nash@cs.ox.ac.uk

**Abstract:** *Decentralized Intelligence Network (DIN)* is a theoretical framework designed to address challenges in AI development, particularly focusing on data fragmentation and siloing issues. It facilitates effective AI training within sovereign data networks by overcoming barriers to accessing diverse data sources, leveraging: 1) personal data stores to ensure data sovereignty, where data remains securely within Participants' control; 2) a scalable federated learning protocol implemented on a public blockchain for decentralized AI training, where only model parameter updates are shared, keeping data within the personal data stores; and 3) a scalable, trustless cryptographic rewards mechanism on a public blockchain to incentivize participation and ensure fair reward distribution through a decentralized auditing protocol. This approach guarantees that no entity can prevent or control access to training data or influence financial benefits, as coordination and reward distribution are managed on the public blockchain with an immutable record. The framework supports effective AI training by allowing Participants to maintain control over their data, benefit financially, and contribute to a decentralized, scalable ecosystem that leverages collective AI to develop beneficial algorithms.

## 1. Introduction

The World Wide Web's evolution from its decentralized origins to today's landscape reflects a complex journey in digital architecture. Originally designed as a distributed network, Web 1.0 envisioned a digital space where data and resources could be shared across multiple nodes without central oversight [1]. However, the emergence of Web 2.0 marked a shift towards centralized platforms, bringing significant efficiency and scalability at the cost of user privacy and control over personal data [2]. While Web 3.0 aims to return to decentralized principles, progress has been gradual [2].

In today's digital landscape, the rapid advancement of artificial intelligence (AI) and the growing volume of data generated across various sectors have created a paradox: while more data than ever is available, much of it remains inaccessible due to fragmentation and siloing within centralized systems. Data is often the lifeblood of AI, yet valuable data remains underutilized due to these silos, where creators and data producers are not fairly compensated for their contributions. This situation limits both data sovereignty and the full potential of AI development.

Personal Data Stores (PDS) have emerged, providing an additional option for enabling data to be held in decentralized peripheries by individuals. Personal Data Stores (PDS) offer a promising solution to address the challenge of data sovereignty and privacy [3], [4]. These systems empower users with decentralized control over their personal data. However, while PDS effectively tackle the issue of individual data control, they introduce a new challenge for AI development. Traditional AI approaches often require re-centralization for data aggregation by third parties, which contradicts the core principles of PDS [3], [4]. The key lies in developing methods that allow AI systems to learn from the rich, diverse data stored across numerous PDSs, without the need to move or copy that data to a central location.

This dichotomy presents two interrelated challenges: ensuring scalable access to data for AI development, and preserving individual data sovereignty in an increasingly data-driven world. We must find better incentives for data contribution and fair distribution of value, addressing how to motivate and reward data providers effectively. The aim of this paper is to design and outline a sovereign decentralized network for AI development that addresses both of these challenges.

Federated Learning (FL) emerges as a promising solution in this context, enabling collaborative model training without requiring data centralization. Bonawitz et al. (2019) made significant strides in their paper "Towards Federated Learning At Scale: Systems Design" addressing scalability issues for mobile Federated Learning (FL) systems. They created a scalable production system based on TensorFlow, designed to handle thousands to billions of users [5]. However, despite these advancements, many systems are still designed to accommodate silos, often mimicking their real-world centralized counterparts. While these systems may decentralize the training process, they often maintain frameworks that favor third-party controllers of data or services. Current FL implementations primarily serve the interests of single-entity providers by focusing on data minimization and reducing data breaches, without fully addressing individual data sovereignty or enabling truly decentralized AI development within sovereign networks.

*DIN* proposes a new approach to decentralized AI with a sovereign FL framework that ensures individual control over data in Personal Data Stores (PDS), enabling learning within sovereign data networks. This framework combines blockchain and IPFS technologies with a sovereign FL architecture, enabling one to adapt tools and workflows from Bonawitz et al. (2019) [5] and others, utilizing TensorFlow (Abadi et al., 2016) [6] and secure aggregation techniques.These adjustments are crucial for enabling scalable and truly decentralized processes within sovereign networks. *Decentralized Intelligence Network* (*DIN*) aims to facilitate decentralized AI development on PDSs, offering scalable access to sovereign-held data while addressing the limitations of current siloed approaches.

*The remainder of this paper is structured as follows:* **Section 2** outlines an exploration of the problem statement and the current limitations in AI and data management. **Section 3** provides an overview of the proposed *DIN* framework systems architecture while **Section 4** presents a theoretical implementation approach of the protocol. Finally, **Section 5 onward** offers its conclusion, outlining future research directions and the broader implications of the work.

## 2. Problem Statement

The current digital ecosystem faces several interconnected challenges:

1. **Data Sovereignty**: Individuals and organizations lack control over their data, often surrendering ownership and usage rights to centralized entities [7].
2. **Limited AI Utilization**: The fragmentation of data across providers and institutions hinders the development of comprehensive, widely beneficial AI models [8].
3. **Access Barriers**: Researchers and developers face significant obstacles in accessing diverse, large-scale datasets necessary for training advanced AI models [9].
4. **Incentive Misalignment**: Current data ecosystems often fail to adequately compensate data providers, discouraging participation in data-access initiatives [10].
5. **Centralization Risks**: Existing AI development paradigms concentrate power and benefits in the hands of a few large tech companies, raising concerns about monopolistic practices and potential misuse of AI technologies [11]. Centralized platforms often create a closed environment with full-stack lock-in and a walled garden, where a single entity decides on value attribution and distribution. This leads to minimal privacy protection and user control, leaving users with limited choices and bargaining power.
6. **Privacy and Security**: Centralized data storage and processing increase vulnerability to breaches and unauthorized access [12].
7. **AI Safety and Control**: The trend towards large, centralized models raises several concerns:
    - Increased risk of developing agent-like behaviors, complicating alignment and control [13].
    - Potential for creating surveillance-like environments due to extensive data access [14].
    - Disproportionate influence of a few entities on global information flow and decision-making [15], [16], [17].
    - Amplification of biases present in training data or introduced by a small group of developers [18].

*Decentralized Intelligence Network* (*DIN*) presents a novel approach to addressing the challenges of data sovereignty, AI development, and privacy in the current digital landscape. By leveraging PDS, federated learning (FL), and blockchain technology, the proposed framework offers a path towards a more equitable, secure, and efficient data ecosystem that respects individual rights while fostering innovation in AI.

### 2.1 Requirements

To address the challenges of collaborative, large-scale AI advancements while preserving data sovereignty and privacy, and ensuring fair, decentralized rewards for participation in federated learning protocols, we establish a framework where data sovereignty of the individual is maintained.

**Hence, we define that data sovereignty of the individual is upheld in this setup, provided that no authority may:**

- 1) Resume access and management controls over the Participants' data, ensuring that control over data remains with the Participants themselves.
- 2) Decide who has access to a Participant's data for federated learning other than the Participants themselves, thus avoiding centralized control over data access.
- 3) Act as a third-party broker to determine which Participant is rewarded for their contributions or the amount of the reward, thereby maintaining fairness and transparency in reward distribution.

This setup ensures that no single authority controls the FL process, preserving participant sovereignty over their data while fostering collaborative AI efforts. Crucially, the framework prevents data other than model updates from needing to leave the PDS, maintaining user privacy and control.

Designed to facilitate a transition towards decentralized, sovereign data stores controlled by individuals, *Decentralized Intelligence Network* (*DIN*) acknowledges that institutional silos and centralized learning are likely to continue. By allowing broad participation in the federated learning (FL) protocol, the *DIN* protocol complements new avenues for access to scalable data for AI engineering in a decentralized fashion. By addressing key issues in the current digital landscape, the protocol aims to create a more fair and secure environment for AI development that respects individual rights and promotes fair participation.

It enhances data access, privacy, and security, encourages standardization of data formats, and enables data monetization for both large and small players. Moreover, by not being limited by geographical constraints of institutional silos, *DIN* ensures a truly global reach and inclusivity, fostering a decentralized and sovereign AI development landscape.

## 3. Background & Related Works

### 3.1 Orchestration

Federated Learning (FL) orchestration processes can be broken down into three main components: 1) Aggregation, 2) Coordination, and 3) Rewards. Each of these components plays a critical role in the orchestration of federated learning, contributing to the overall effectiveness and efficiency of exploring FL frameworks. In this context, we assume a discussion surrounding FL orchestration specifically for work on decentralized solutions, while acknowledging that personal data stores (PDS), which are crucial for preserving data sovereignty, are not elaborated on here for brevity. For more details, see references [3], [19], [20], [21].

### 3.1.1 Aggregation

Aggregation is a fundamental component of FL that involves combining local model updates from multiple participants to form a global model [22]. Traditionally, aggregation is handled by a central server that collects, processes, and averages these updates.

According to this paper's definition of sovereignty, as long as the coordination and rewards processes are trustless and decentralized, the aggregation step itself does not need to be decentralized. Thus, preserving the ability to use scalable tools and workflows similar to those described by Bonawitz et al. (2019)—which build on their earlier work (2017) that detailed Shamir's secret sharing as a component of secure aggregation—is feasible [5], [23]. Bonawitz et al.'s (2019) system addresses scalability while incorporating privacy-preserving techniques such as the secure aggregation protocol from their 2017 work [5]. While a detailed exploration of additional privacy-enhancing techniques, like differential privacy (McMahan et al., 2018), is beyond this paper's scope, these methods illustrate how colluding agents cannot infer information about a remaining agent when $N \geq 3$. Algorithm developers can tailor the choice of privacy techniques and protocols within the *Decentralized Intelligence Network* (*DIN*) to their specific requirements [5].

However, exploring alternative protocols that could offer interchangeable aspects for the *DIN* framework remains of interest. While the current requirements do not necessitate an in-depth exploration of these alternatives to achieve the frameworks objectives, recent research has examined decentralized aggregation methods that eliminate the need for a central server. These methods aim to enhance scalability, reduce reliance on centralized entities, and improve data privacy—areas still worth exploring.

One such proposal, IPLS, introduced by et al. (2021), collectively trains a model in a peer-to-peer fashion without the assistance of a server by using an IPFS-based protocol [24]. Unlike the centralized setting, where only the server is responsible for storing, updating, and broadcasting the model to the participating agents, IPLS splits the model into multiple partitions replicated on multiple agents [24]. However, this framework requires extensive expertise to handle various model types and compression techniques, making it difficult to train more complex algorithms.

Vincent et al. (2020) proposed "Blockchain Assisted Federated Learning" (BC-FL), which replaces the need for a central server in the aggregation process by leveraging a public blockchain [25]. This approach considers that local model updates can be received by miners through a gossip protocol over the P2P network [25]. However, gossip-like protocols are notorious for diverging from the real value and failing to reach consensus [26].

Ramanan et al. (2020) proposed "BAFFLE," an aggregator-free FL protocol that eliminates the need for a central server during the FL process [27]. However, this requires splitting and compressing machine learning models on the blockchain itself, posing significant challenges due to the complexity of model compression techniques and the extensive research needed to make this feasible.

Overall, the framework could benefit from exploring these possibilities and adapting emerging technologies to effectively address privacy and scalability concerns.

### 3.1.2 Coordination

Coordination in Federated Learning (FL) traditionally involves a central authority managing participant interactions and model updates. Bonawitz et al. (2019) demonstrated that a centralized approach can be both scalable and secure, with their system architecture of Coordinator, Master Aggregators, and subgroup of Aggregators, thereby handling up to 10,000 active devices simultaneously and potentially scaling to billions [5].

However, centralized coordination poses potential problems such as dishonest aggregation, network failures, external attacks, and ensuring protocol adherence [28]. To address these issues and maintain data sovereignty, *Decentralized Intelligence Network* (*DIN*) proposes using blockchain technology for coordination.

**Blockchain offers several advantages for FL coordination:**

1. Decentralization: Prevents any single authority from controlling data access, preserving individual sovereignty [22] .
2. Transparency: An immutable ledger records and verifies updates, enhancing trust among participants [21].
3. Fault tolerance: The peer-to-peer design improves system integrity [29], [30].
4. Computational benefits: Enhances round delineation, model selection, and model aggregation in a decentralized manner [31].

*DIN* incorporates a hierarchical aggregation structure similar to Bonawitz et al. (2019), but replaces centralized coordination with a protocol distributed on a public blockchain smart contract [5]. This approach maintains scalability while ensuring open access to FL protocols. The coordination process utilizes a secure aggregator server with a blockchain node and a public on-chain smart contract as detailed in **Sections 4 and 5**.

Our framework builds upon the Secure Aggregation principle introduced by Bonawitz et al. (2017) [23], which uses encryption to make individual devices' updates uninspectable by the server. To overcome the quadratic computational costs of secure aggregation while still allowing large numbers of participants, the system runs an instance of secure aggregation on each Aggregator actor [5]. These Aggregators produce intermediate sums from groups of at least k devices, where k is a parameter defined by the FL task. The Master Aggregator then combines these intermediate results into a final aggregate for the round without using Secure Aggregation [5].

While blockchain can introduce some network delays [27], the benefits of sovereignty and decentralization outweigh this drawback in practical applications. In reality, these network delays may be acceptable depending on the specific use case. Importantly, *DIN* uses a public blockchain to prevent re-centralization of data and overcome competitive interests of institutional stakeholders. This is crucial, as even private blockchains can lead to centralization if a single authority orchestrates FL protocols and selects participants [10].

By leveraging blockchain technology, *DIN* aims to address the limitations of centralized coordination while maintaining the scalability benefits demonstrated by Bonawitz et al. (2019). This approach allows for potentially handling large numbers of participants efficiently while ensuring data sovereignty and transparent, decentralized coordination. This framework can accommodate highly scalable federated learning involving large numbers of Participants, as per Bonawitz et al (2019)., but in a decentralized manner [5], enabling wider application of these tools and workflows within sovereign networks.

### 3.1.3 Rewards

The reward mechanism in federated learning (FL) incentivizes participants to contribute their computational resources and data, binding decentralized participation and enabling new use cases. Addressing the need for an incentive mechanism is essential for quantitative and evaluative inquiries in FL [28]. The reward mechanism in federated learning (FL) incentivizes participants to contribute their computational resources and data, binding decentralized participation and enabling new use cases. Addressing the need for an incentive mechanism is essential for quantitative and evaluative inquiries in FL [28].

A well-designed reward system is crucial for boosting engagement and ensuring fair compensation. In siloed systems, data generators often remain unrewarded despite being key stakeholders. However, **decentralized reward mechanisms** using smart contracts and blockchain can address these issues by enhancing transparency, fairness, and resistance to manipulation. Smart contracts (SC) can orchestrate multiple FL tasks simultaneously across different sets of devices, ensuring that contributions are fairly evaluated and compensated [22]. In contrast, it is worth noting that private blockchains rely on a trusted setup, where the orchestrator might collude with the Model Owner of or stakeholders potentially acting maliciously. Traditional FL lacks incentives to encourage clients to follow the protocol honestly and provide reliable data [32].

Hence, several prior works emerged, such as 2CP by Cai et al. (2020), and Blockflow by Mugunthan et al. (2020), which have outlined procedures for measuring Participant contributions. 2CP employs Substra for step-by-step evaluation [31], [33], while Blockflow evaluates overall scores based on the median score reported for each model and the inverse of the maximum difference between reported and median scores [34]. However, these frameworks are limited to small numbers of participants, as they were designed to mimic their real-world centralized counterparts. They do not address the scalability required for larger participant numbers or the associated security guarantees [33], [31], [34]. Both BlockFlow (2020) and 2CP (2020) implemented a 1:1 ratio of evaluators to participants, with each participant evaluating every other participant's score [31, p. 2], [33]. For instance, BlockFlow (2020) demonstrated an average absolute difference of less than 0.67% between evaluators' scores across various limited numbers of agents (1, 25, 50, and 100) using income data. However, their experiments were constrained to these participant numbers and did not explore larger scales [31]. Furthermore, both of these frameworks assume all Participants must act as Evaluators in the rewards process, which is not scalable as costs rise asymptotically with the number of Participants [31]. For example, with 100 Participants, all would need to download and evaluate the models of the other N - 1 Participants.

Unlike existing works such as BlockFlow (2020) and 2CP (2020), which do not fully address scalability in reward distribution, the *DIN* framework is designed to ensure rewards are issued in a scalable manner while meeting sovereignty requirements. The proposed architecture utilizes a public blockchain to orchestrate a 'trustless' process for coordination and rewards, eliminating the need for a third party [22], [32].Scalability in the aggregation process is managed through on-chain coordination by *DIN* smart contracts (SC). These SCs handle various tasks, such as managing aggregators, guiding protocol interactions, and ensuring coherent operation. This approach follows the method outlined by Bonawitz et al. (2019), which involves deploying multiple aggregators in each round. This approach ensures scalability by handling multiple aggregators simultaneously. Additionally, for the reasons discussed above, the constraints of the rewards process will influence the size of these aggregator subgroups in the federated learning (FL) process, similar to the architectures employed by Bonawitz et al. (2019).

*DIN* framework integrates two key contributions to enhance scalability and efficiency:

1. **Role Delineation**: The framework introduces a clear separation of roles by delineating Participants into two distinct categories: Participant and Evaluator. In this setup, the Evaluator is specifically assigned the task of evaluation, while the Participant primarily acts as a Personal Data Store (PDS) holder. This role separation allows for task specialization and ensures that not all Participants are required to participate in evaluations, facilitating scalability.
2. **Evaluator-to-Participant Ratio**: Inspired by BlockFlow's (2020) recommendations, the DIN framework integrates a ratio of Evaluators to Participants to perform evaluations. Evaluators are randomly selected ($Q \ll N$) to assess Participants' work within each subgroup FL aggregator group each round. For example, in an aggregator group with 100 Participants, 10 Evaluators might be assigned. While a sufficiently large number of Evaluators is theoretically expected to improve accuracy and resist potential manipulation, the exact effectiveness of this ratio in ensuring accurate results and maintaining resilience against a majority of malicious agents ($M < N/2$) will need to be validated through empirical testing [31]. This integration supports scaling across increasing FL rounds, ensuring both robustness and efficiency [31].

These integrated aspects collectively address the challenges of scalability and task specialization in the federated learning process, enhancing the overall effectiveness of the *DIN* framework.

However, to ensure generalizability and maintain the integrity of evaluations due to the delineation of roles, the Model Owner must publish a control dataset for Evaluators to benchmark Participant scores. Previously, Participants used local datasets for this purpose.Previously, Participants used local datasets for this purpose. However, a new risk is introduced with benchmarking using a control dataset issued by the Model Owner, including the possibility of colluding parties using the control dataset to strategically enhance their model performance or manipulate evaluation outcomes to gain undeserved rewards.

Therefore, control datasets must be securely transmitted to the secure server with a blockchain node, with their location communicated to proven (i.e., proof-of-stake) Evaluators on-chain. Additionally, a Trusted Execution Environment (TEE) may be employed to further protect the dataset. These procedures are elaborated in **Section 5.2, Decentralized Auditing Protocol.**

Evaluators will then apply contributivity scoring procedures, such as 2CP's Substra or Blockflow's median scoring, to assess performance accurately [31], [33]. To enhance security and privacy further, the auditing process can employ Zero-knowledge proofs (ZK-proofs), allowing participants to prove the validity of their computations without revealing underlying data [35], [36].

This decentralized approach, executed on a public blockchain consensus, prevents any single entity from manipulating the reward distribution, thereby maintaining trust in the system. Furthermore, future experiments should assess whether the score allocation differences persist when using a larger ratio of Participants to Evaluators, as this may depend on the data type being trained upon.

Additionally, validating the protocol across heterogeneous data sources is crucial. These subgroups should undergo rigorous stress testing and experimentation to confirm their effectiveness and scalability. Moreover, while this paper assumes a steady-state system with constant numbers of Participants and Evaluators per aggregator subgroup FL round, addressing the dynamic nature of networks is essential for future research.

# 4 Proposed Solution: Systems Architecture and Overview

*Decentralized Intelligence Network* (*DIN*) addresses critical issues by providing a theoretical framework and practical implementation for a decentralized, sovereign data ecosystem. This approach enables scalable AI development while preserving individual rights and promoting fair participation. Importantly, *DIN* focuses on smaller, distributed models designed to run on consumer hardware, aligning with a safety-conscious and privacy-preserving approach to AI development.

The *DIN* protocol employs smart contracts (SC) to manage key processes, including for coordinating and rewarding AI training, and for enabling Evaluators to assess participant contributions in a secure, novel proof-of-stake ecosystem. By leveraging these decentralized mechanisms, the framework ensures scalable, sovereign data solutions that respect individual rights and drive technological advancements.

**This framework's design adheres to the requirements detailed in Section 2 to uphold data sovereignty.** By ensuring that no single authority controls the federated learning process, this approach preserves participant sovereignty over their data and

Page 6

promotes collaborative AI efforts. Additionally, the framework ensures that only model updates, and not raw data, are transferred out of Personal Data Stores (PDS), thereby maintaining user privacy and control.

To address these challenges, this paper proposes a comprehensive framework that integrates federated learning (FL) and a trustless rewards mechanism. This research focuses on enabling collaborative, large-scale AI advancements while preserving data sovereignty and privacy, and ensuring fair, decentralized rewards for participation in federated learning protocols.

A critical component of the frameworks approach is ensuring that federated learning protocols can effectively leverage data from numerous Personal Data Stores (PDS) while maintaining individual data sovereignty. As the number of PDS participants increases, federated learning protocols must scale accordingly, presenting new challenges in designing systems that can accommodate a larger and more diverse network of contributors.

Equally important is the implementation of a scalable and decentralized reward system that aligns with the principles of federated learning and PDS. This system must acknowledge the contributions of participants who allow their local data to be used for training models and provide computational resources, while maintaining transparency and fairness without centralized control.

**The proposed framework consists of three key elements:**

- 1) Personal data stores (PDS) to ensure individual data ownership.
- 2) A scalable federated learning (FL) protocol on a public blockchain for decentralized AI training.
- 3) A trustless rewards system to incentivize participation and ensure fair reward distribution.

This setup ensures that no single authority controls the FL process, preserving participant sovereignty over their data while fostering collaborative AI efforts. Crucially, the framework prevents data other than model updates from needing to leave the PDS, maintaining user privacy and control.

The proposed *DIN* protocol is designed to facilitate a transition towards decentralized, sovereign data stores controlled by individuals. It acknowledges that institutional silos may continue to participate as single-identity entities, potentially leading to a period where these system architectures co-exist. This approach does not preclude institutions from contributing to the FL protocol, and the *DIN* protocol may act as a catalyst for this transition.

By addressing these issues, the proposed framework aims to create a more equitable and secure digital ecosystem, enabling decentralized, sovereign AI development that respects individual rights and promotes fair participation. It enhances data access, privacy, and security, promotes the standardization of data formats across systems, and encourages data monetization for both large and small players. Additionally, by not being limited by geographical constraints of institutional silos, the *DIN* protocol ensures a truly global reach and inclusivity.

By prioritizing models that can operate on consumer devices, *DIN* aims to avoid a privacy and centralized-control dystopia where AI relies solely on centralized servers. In such a scenario, server operators could monitor all actions and shape AI outputs according to their biases in ways participants cannot escape. Instead, *DIN*'s approach empowers individual users, allowing them to leverage AI capabilities while maintaining control over their data and reducing reliance on centralized infrastructure.

**Key participants in this system include:**

- **Participants:** Individuals who own and control their data stores, contributing data to the FL process while maintaining privacy and benefiting from collaborative AI training.
- **Model Owners:** Entities such as companies or researchers that utilize the FL protocols to enhance their models with decentralized data, without compromising individual data sovereignty.
- **Evaluators:** Network-staked entities responsible for decentralized auditing, ensuring transparency and fairness in evaluating participant contributions and distributing rewards.

This decentralization-focused strategy contrasts with acceleration-focused approaches that emphasize ever-larger models and computing clusters. *DIN* prioritizes a vision of AI development that mitigates privacy risks and centralized control associated with large, server-based models. By leveraging consumer hardware, this approach aims to create AI systems that are more like bounded tools than independent agents, potentially offering significant advantages in terms of AI safety, control, and individual empowerment.

In summary, this paper presents a novel approach to addressing the challenges of data sovereignty, AI development, and privacy in the current digital landscape. By leveraging PDS, federated learning, and blockchain technology, the proposed framework offers a path towards a more equitable, secure, and efficient data ecosystem that respects individual rights while fostering innovation in AI.

## 5. Methodology

**5.1** *Decentralized Intelligence Network* (*DIN*) **Protocol**

*Decentralized Intelligence Network's* (*DIN*) protocol operationalizes the federated learning (FL) architecture outlined in **Section 4**. Built on a decentralized public blockchain infrastructure, the *DIN* protocol manages the coordination and rewards process for training machine learning models using data stored in Participant-owned Personal Data Stores (PDSs). This approach ensures data sovereignty while enabling collaborative AI development.

Participants opt into FL protocols defined by smart contracts (SC) on a public blockchain. The protocol leverages the blockchain to coordinate the FL process and manage rewards, while ensuring that raw data remains within the Participant's sovereign datastore. Only model updates are shared during the FL process, preserving privacy and control.

To enhance scalability and computational efficiency, the DIN protocol incorporates an off-chain decentralized file storage system, such as the InterPlanetary File System (IPFS) [37]. This system provides a location for uploading and downloading model updates during the learning process, optimizing participation costs and complementing the blockchain's transaction recording capabilities.

This setup ensures a fair and transparent reward system while maintaining data sovereignty and reducing reliance on centralized infrastructure. The following sections detail the specific methodologies and operational mechanisms of the *DIN* protocol, including the roles of key participants such as Model Owners, Participants, and Evaluators.

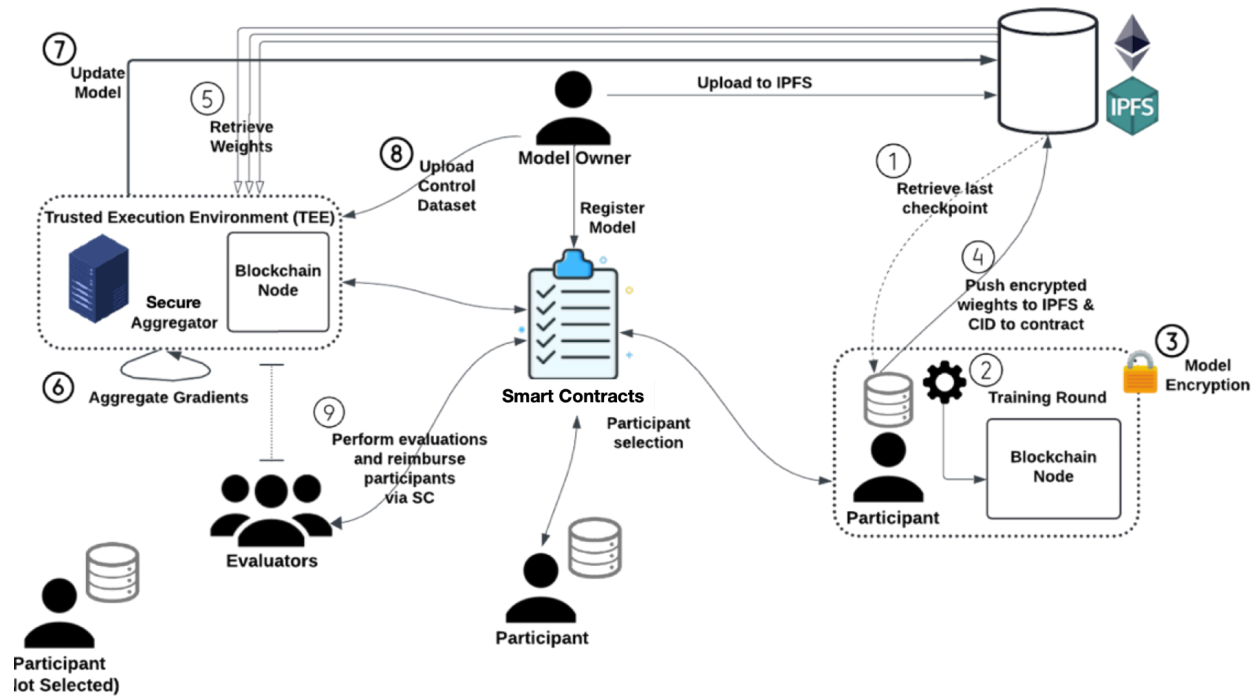

***Figure 1.*** *Global overview of a training round. Adapted from Consensys Health [10].*

1. **Model Owner:** a. Deploys *intelligence* smart contracts (SC) on the blockchain. Although referred to as "*intelligence*," multiple contracts may work in unison on-chain, fulfilling different roles to enhance scalability and efficiency (e.g., *DIN* protocol, evaluator registry, evaluator staking, aggregator management, NFT staking for evaluators, reward distribution, etc.). b. Creates genesis model, uploads to IPFS, records its CID on the contract. c. Deposits reward amount in smart contract (SC) to be allocated to Participants after FL rounds.
2. **Upon Model Owner's transaction confirmation:** a. Participants can see the genesis model CID and download it from IPFS. b. Using their own data stores, Participants run training iterations on model. c. Participants encrypt their models using secure aggregation protocol (e.g. Bonawitz et al. 2019) [5]. d. Participants upload their encrypted updated models to IPFS, recording CIDs on *intelligence SC*. e. This completes one FL training round; Participants wait for the next round.
3. **Aggregation:** a. The secure aggregator server, operated by the Model Owner, fetches all encrypted weights from decentralized storage as instructed and verified on-chain. b. Standard Process: The Model Owner uses the secure aggregator to perform aggregation using secure aggregation protocol c. The secure aggregator server, using a blockchain node operated by the Model Owner, fetches encrypted model weights from IPFS via CIDs recorded on the blockchain, ensuring secure processing. d. To manage scalability, a master aggregator spawns multiple aggregators each round, as in Bonawitz et al. (2019) [5], with device management integrated into on-chain coordination by the *DIN* smart contracts (SC). The protocol's existing components handle device connections, make local decisions based on instructions from the Coordinator on-chain, and forward devices to the aggregators, ensuring efficient allocation while minimizing communication with the on-chain Coordinator. e. The aggregated results are confirmed on-chain before the global model update is revealed to the Model Owner. f. The Aggregator uploads the new global model update to shared storage and updates on-chain *intelligence* SC with the new pointer. g. Multiple secure aggregator servers or the preferred Model Owner environment for aggregation can be instantiated to handle larger workloads or enhance decentralization and reliability.
4. **Next Training Round:** a. A new global model is published by the Model Owner for the next training round. b. **Optional:** Participants can check all CIDs of the previous round's model updates in their aggregator subgroup. b. Download updates from IPFS and independently calculate the mean aggregate for their subgroup (all participants in a subgroup reach the same result). c. Each subgroup performs the average of their own subgroup's global model and then averages other aggregator groups' averages as published on IPFS located on-chain to reach global model published by Model Owner.
5. **Model Aggregation Continuation:** a. The Model Owner continues to aggregate FL round updates, averaging their updates as rounds progress. b. The Model Owner tests the average global model update against their control dataset to determine when they are satisfied with the training. c. The decision to end or continue training is communicated to the SCs either in advance or during training, depending on the Model Owner's availability of funds. d. The Model Owner signals the final round, after which one additional round occurs to evaluate Participants' rewards before the final global model is revealed to the Model Owner.
6. **Post-Training:** a. The Model Owner encrypts the control dataset and uploads it to the aggregation server and may open a Trusted Execution Environment (TEE) and/or utilize privacy-preserving techniques if desired. b. The *intelligence* SC randomly assigns a standardized ratio of Evaluators to Participants (e.g., 1:10) according to the Model Owner's needs [31]. c. Evaluators, who are randomly assigned to a specific subgroup FL aggregation round for scalability, benchmark all Participant models of that aggregator group inside the secure aggregator server, potentially using TEE and/or other privacy-preserving techniques for off-chain benchmarking (unless there is a disagreement, in which case benchmarking is conducted on-chain).
7. **Evaluators:** a. Anyone staking native token can be an Evaluator. b. Evaluate Participants' models against the control dataset in the secure server environment. c. Submit evaluation consensus scores with ZK-proofs to *intelligence* SC (see **2.3.1 Decentralized Auditing Protocol** for rewards).
8. **Distribute Rewards:** a. *Intelligence* SC calculates reward fraction for each participant based on objective scores (e.g., as Shapley Values, Substra Scoring, Median Scoring, and other open-preference scoring methods) [31], [33], [38], [39], [40]. b. Distributes Model Owner's deposited reward accordingly, per recorded scores.

In the related works section, we briefly reference the protocol by Bonawitz et al. (2019), which employs synchronous rounds with subsets of devices for scalable federated learning. Their Secure Aggregation method ensures privacy by encrypting device updates, and intermediate results are aggregated by dedicated actors to manage computational costs. This framework builds on these principles, enhancing scalability and privacy while addressing sovereignty and decentralized participation. By integrating subgroup aggregation and subset evaluation, we extend Bonawitz et al.'s approach to support sovereign networks and broader applications [41], [42].

To incentivize participation in this scalable and decentralized FL process, we propose integrating a scalable, "trustless" rewards mechanism—one that does not require a third party for transactions. This mechanism is discussed in the following section.

## 5.2 Decentralized Auditing Protocol

Delineating the roles of Participant and Evaluator in the protocol raises concerns about the potential misuse of the control dataset by Evaluators in federated learning (FL) scenarios. In previous examples, each Participant evaluated every other Participant using their data and an objective scoring metric [31], [33]. However, once we distinguish between Participant and Evaluator and adjust the protocol to scale and generalize to other data types, a new issue emerges. Evaluators might download and illicitly share the control dataset published by the Model Owner with Participants in one or more aggregator subgroups in a FL rounds when benchmarking Participants' contributions after Model Owenr signals final round, potentially leading to harmful activities such as model poisoning or unfair compensation during the model training process.

To mitigate these risks, implementing secure evaluation mechanisms where the test dataset remains concealed from the Evaluators is essential. Evaluators can prove to the system they correctly evaluated against the control dataset without accessing it, mitigating risks of misuse or leakage. Evaluators can be provided with high-quality, well-distributed, and highly representative control datasets by the Model Owner. Evaluators can use this as a benchmark to evaluate each Participant's models as shown in **Figure 1**. The step-by-step protocol elaborates on the processes of Evaluators' involvement in the **decentralized auditing protocol** within the rewards process, as illustrated in **Figure 1**, as follows:

1. Model Owner selects secure computation method: a. Opens a TEE within the aggregation server, and/or b. Chooses alternative privacy-preserving computation meeting security standardsModel Owner distributes encrypted control dataset to chosen secure environment.
2. Within a secure environment: a. Evaluators compute performance metrics (such as accuracy, precision, and recall) on models using a control dataset and benchmark these metrics using objective scoring methods (e.g., Shapley Values of Distribution, Substra, Median Scoring, etc) [31], [33], [38], [39], [40]. b. Evaluators utilize remote attestation mechanisms to prove secure execution. c. Evaluators generate Zero-Knowledge Proofs (ZKPs) for each metric.
3. Contingency for MPC: a. Participants encrypt model updates using MPC or other privacy-preserving technique to benchmark Participants' scores. b. Evaluator evaluates models on encrypted test dataset using privacy-preserving protocol.
4. Evaluators submit ZKPs and encrypted evaluations to blockchain *intelligence* SC.
5. *Intelligence* SCs verifies ZKPs using a consensus mechanism (e.g., majority agreement).
6. Based on verified scores, *intelligence* SC calculates and distributes rewards transparently.
7. All privacy-preserved transactions are recorded on the blockchain for an immutable audit trail.
8. Privacy-preserving techniques (e.g., MPC) along with TEEs and ZKPs, provide multiple layers of security. MPC, in particular, can conceal data, TEEs isolate it, and ZKPs verify correctness without revealing information.
9. Protocol protects against insider threats by computing within secure environments and exporting only ZKPs.
10. Final Global Model Update: a. Model Owner signals end of training process upon satisfaction with model performance. b. Final Global Model update revealed to Model Owner after completion of rewards process.

*DIN* can utilize any contributivity scoring procedure for Evaluators to perform their off-chain evaluations of Participants' contributions. The specific procedure depends on the context of the learning task being conducted. This evaluation process is triggered when the Model Owner is satisfied with the global model's performance metrics (e.g., F1 score, accuracy, etc.), and occurs after a given number of FL rounds have been iterated in the FL process.

Following the approach of Bonawitz et al. (2019), this protocol can employ synchronous rounds with subsets of devices [5]. Evaluators are randomly assigned to aggregator Federated Learning (FL) participant subgroups at a specific ratio (e.g., one Evaluator for every ten Participants). This approach enables high scalability, allowing the system to handle a large number of devices while maintaining efficiency.

Evaluators perform their off-chain evaluations of each Participant's model using the control dataset published by the Model Owner, encrypt their score, and report their encrypted scores with a Zero-Knowledge Proof (ZKP) to the *intelligence* SC. Each Evaluator first reports to the smart contract the set of Participants whose models were successfully validated (i.e., within an acceptable bound specified by the Model Owner). Participants who fail this test are eliminated in that particular FL round. Once all the scores are received, each Evaluator provides the decryption key to provably reveal their score to the *intelligence* SC.

This protocol ensures secure and reliable model evaluations within a secure server environment and can incorporate privacy-preserving techniques such as Multi-Party Computation (MPC), as well as third-party Trusted Execution Environments (TEEs) selected by the Model Owner, safeguarded by ZKPs and blockchain technology. These precautions ensure that the protocol uses encryption to make individual evaluations uninspectable by any central authority or even other participants in the FL process, as detailed in the threat model (**see Section 6**). These techniques could be applied in the context of assessing rewards, similar to how they're used in aggregation.

The protocol aligns with the Model Owner's incentives, who, despite bearing the cost, benefit from robust data contributions and trustworthy evaluation processes necessary for successful model training and improvement. Crucially, it preserves data sovereignty for Participants—no central authority determines access to their data, which never leaves its original storage. Rewards are determined in a decentralized manner by a public blockchain smart contract based on pre-defined, auditable, and transparent metrics, eliminating the need for a central authority to decide compensation. This approach overcomes the potential issues of centralized systems, such as dishonest aggregation or external attacks, while maintaining the scalability benefits demonstrated by Bonawitz et al. (2019) [5].

This decentralized auditing protocol maintains Participants' autonomy while enabling secure, reliable, and incentive-aligned model evaluations. It addresses the challenges of sovereignty and decentralized participation within a novel framework, enabling wider application of these tools and workflows within sovereign networks.

## 6. Threat Model

The papers threat model addresses potential risks in the federated learning (FL) process, ensuring robust security and privacy. The protocol is resilient to up to 50% malicious participants, leveraging public/private key cryptography and a proof-of-stake consensus mechanism. By using immutable storage on IPFS we ensure data integrity. Additionally, the use of Zero-Knowledge Proofs (ZKPs) and Trusted Execution Environments (TEEs) mitigates risks associated with model evaluation and reward distribution. This comprehensive approach ensures the security and reliability of the FL process, maintaining participant trust and data sovereignty.

Firstly, in an experiment with N agents, it is resistant up to $M \in [0, N/2)$ agents neglecting to follow the protocol for the experiment to maintain its integrity [31]. For example, public/private key cryptography and a proof-of-stake consensus protocol secure the Ethereum blockchain. Currently, there are no feasible attacks on the Ethereum Network, without controlling 50% of the computational power of the entire Ethereum network and such an attack has never been successful on the Ethereum mainnet [43].

Secondly, as a public blockchain is public and anonymous, clients could enroll multiple times in an experiment and thus have a disproportionate participation. However, through decentralized identity verification, verifiable credentialing, or manual processes, agents can ensure that each other agent controls only one account [19], [20], [44].

Third, IPFS is immutable, meaning agents cannot change their model after submitting the cryptographic hash to the smart contract [37]. Like in BlockFlow, the *DIN* protocol requires each agent to report if it can load strictly more than N/2 models, and have strictly more than N/2 agents report the same for their model. The *DIN* threat model guarantees that there are strictly more than N/2 honest Participants. Additionally, as long as N/2 or more Evaluators who receive these models for evaluation are honest, which the *DIN* protocol guarantees, the system remains resistant to N/2 attacks. Since IPFS allows anyone to share any content, one or more honest parties would share the model with all other Participants if they are unable to retrieve a model directly from the source (e.g., due to firewall restrictions). Therefore, each Participant would still be able to obtain all necessary models [31], [37].

Fourth, there are several possible attacks on the contribution scoring procedure itself. Malicious models are those with weights that do not reflect a truthful dataset, such as models trained on randomly generated data or inverted output features. Naively averaging such models into a global model would likely harm the shared objective. The *DIN* protocol can choose contribution-scoring procedures that penalize those who submit malicious models. For instance, BlockFlow (2020) uses a contributivity score system where lower scores result in less cryptocurrency received [31], [37]. In this system, any agent with an evaluation more than 0.5 away from the median score receives an overall score of 0 and no share of the cryptocurrency pool [31], [37]. This penalizes attempts to fabricate scores, as the protocol limits a Participant's overall score to the evaluation furthest from the median [31], [37].

Fifth, Participants can collude during the training process to submit better models by secretly sharing raw data or models among $M<N2$ colluding Participants [31], [37]. The *DIN* protocol rewards Participants who contribute strong models, and it is acceptable for multiple Participants to submit identical models. Such collusion is not considered an attack, as it is similar to having many Participants with strong datasets [31], [37]. For attacks by Evaluators in the evaluation process, the I smart contract can use encryption and a commit-then-reveal protocol (e.g., Secret Sharing MPC, Elliptic Curve Diffie-Hellman keys, etc.) to prevent Evaluators from copying others' scores without collusion [45]. If a minority subset of malicious Evaluators reports perfect 1.0 scores for certain models and 0.0 scores for all others (e.g., models from honest agents), the median score is guaranteed to be between the minimum and maximum scores reported by the honest agents, as long as there are strictly fewer than half malicious Evaluators [31], [37]. Evaluators are incentivized to stake an NFT (non-fungible token) to gain the right to evaluate participant models in the rewards process within a proof-of-stake (PoS) ecosystem. This staking mechanism involves the use of a native NFT token standard. The system operates as a self-assessed value framework, incorporating Harberger taxation, proceeds of which fund public good systems. Evaluators found acting maliciously are slashed from the network, losing some or all of their stake, thus maintaining network security and incentivizing honest work.

Sixth, in this papers threat model, it is crucial that the control dataset provided by the Model Owner remains encrypted to prevent its misuse. If the control dataset were accessible to colluding Participants, Model Owners, Evaluators, or other entities, they could exploit it to skew the reward distribution. For example, colluding parties could use the control dataset to strategically improve their model performance or manipulate evaluation outcomes to gain undeserved rewards. Encrypting the control dataset ensures that it cannot be revealed or utilized by these entities to unfairly influence the results. To enhance security further, dual protection strategies can be employed. For instance, the preferred scoring method, such as median scoring used by BlockFlow (2020), can be integrated into the protocol [31]. In this approach, any score deviating significantly from the median—beyond a specified threshold—can be penalized. BlockFlow's method maps any score differing by more than 0.5 from the model's median to a score of 0, with an a priori score set at 0.5 [31]. This mechanism encourages evaluators to provide honest assessments by penalizing scores that deviate substantially from the median. This method helps mitigate the risk of anomalous scores due to collusion and maintains fairness in the reward distribution process. Overall, encrypting the control dataset and employing robust scoring mechanisms collectively safeguard the integrity of the evaluation process and prevent potential manipulation by malicious actors.

Seventh, both the aggregation process involving trusted third-party secure aggregation servers and other such as Trusted Execution Environments (TEEs) with Zero-Knowledge Proofs (ZKPs) introduce distinct threat models. Concerns with aggregation hardware include potential data interception and manipulation, insider threats at third-party providers, hardware vulnerabilities such as side-channel attacks, and compliance issues with data protection regulations [46], [47]. TEEs, while isolating sensitive computations, face risks from hardware exploits, software vulnerabilities, and third-party trust issues. Additionally, implementations of ZKPs must be carefully managed to avoid cryptographic flaws that could undermine their effectiveness [48]. To counteract these threats, robust encryption, rigorous access controls, regular security audits, and compliance assurance are employed [48]. These measures ensure that data remains confidential and integral, reducing the dependency on trust by making processes transparent and verifiable through smart contracts on the blockchain. This strategy enhances security and stabilizes residual risks within a robust, transparent operational framework.

## 7. DIN Applications

*Decentralized Intelligence Network* (*DIN*) offers a scalable and versatile framework for learning from sovereign, individually owned data stores, supported by a reward system designed to boost participation. This paper lays the groundwork for future research on decentralized services, aiming to leverage sovereign data stores for innovative algorithm development.

**Key use cases include:**

- **Healthcare**: Patients store their health data in self-sovereign data stores, controlling access and sharing model updates securely. Medical researchers and healthcare providers can access the FL protocol on-chain to train AI models on this data, improving diagnostics and treatment plans without ever seeing the raw data. Patients can be financially rewarded for contributing to medical research, and they can use these rewards to help cover insurance premiums, thereby lowering the barrier to providing accessible healthcare.

- **Finance**: Individuals store their financial transaction data in decentralized data stores. Financial institutions can access the FL protocol on-chain to provide personalized financial advice and develop new financial products based on aggregated insights. Users remain in control of their data and can receive rewards for their participation, fostering a transparent and incentive-aligned financial ecosystem.

- **Education**: Students store their academic records and learning progress in self-sovereign data stores. Educational institutions can access the FL protocol on-chain to tailor learning experiences and provide personalized support without accessing the raw data. Students can receive incentives for allowing their data to contribute to educational research and improvements, funding some of their education costs.

- **Smart Cities:** Residents store data related to their energy consumption, transportation patterns, and other smart city metrics in self-sovereign data stores. City planners and utility providers can access the FL protocol on-chain to optimize city services and infrastructure without accessing the raw data. Individuals receive rewards for allowing their data to promote sustainable and efficient urban living, and they can use these rewards to contribute toward various living costs, thereby improving their quality of life.

- **Agriculture:** Farmers store data on crop yields, soil conditions, and weather patterns in decentralized data stores. Agricultural researchers and companies can access the FL protocol on-chain to develop better farming practices and technologies. Farmers retain control over their data and receive rewards for their contributions, fostering innovation and sustainable agriculture. Additionally, farmers can use the rewards they receive to contribute to crop insurance, benefiting from monetizing the data they collect and reinvesting it into their local ecosystems.

These use cases emphasize the use of decentralized data management and federated learning protocols to ensure privacy while allowing industries to leverage valuable insights for enhancing their services.

## 8. Tokenomics, Governance, and Public Goods

This section introduces a novel protocol designed to incentivize and manage participation within a *Decentralized Intelligence Network (DIN)*. It highlights the significance of integrating economic and social dimensions into modern systems architecture, emphasizing the importance of tangible incentives and broader systems design. This approach aligns with the perspectives of early internet pioneers and recent literature [49].

The protocol incorporates a novel public goods funding mechanism that not only secures the network but also integrates seamlessly with existing public goods ecosystems. It leverages NFT staking, unique evaluation mechanisms, and principles of Partial Common Ownership (PCO) to foster a circular economy for the development and enhancement of global AI models. While the staking mechanism is central to the protocol, it does not necessarily require a native token or coin, though the potential for such integration could be explored.

**NFT Proof of Stake (PoS) Mechanism for Evaluators:** This component explores the novel application of combining innovative evaluation mechanisms and staking methods with Harberger taxation and PCO principles. It focuses on using NFT staking to secure the network and aims to establish standard, open-source implementations of Partial Common Ownership (PCO) of Ethereum ERC721 NFTs [50], [51]. By integrating these methods with Harberger taxation, this approach not only strengthens network security but also seeks to contribute revenue to public goods ecosystems, potentially enabling self-funding development of the *DIN* and other adjacent public goods ecosystems [51]. Core components of the staking mechanism are detailed:

- **Network Fees**

- The Model Owner, who trains the algorithm and pays Participants for the Federated Learning (FL) process on sovereign data stores, is assumed to pay this fee as part of the process.
- **The fee distribution from rewards to participants and evaluators is dynamic and remains an open question.** It could be either an added tax on the reward or a distribution model, such as allocating 97% of the reward to participants and 3% to evaluators.
- This fee is separate from and in addition to any blockchain-specific gas costs, and is a part of the estimated costs of the rewards process.
- The exact calculation and distribution method for this reward is an open question for experimentation.

- **Fee and Reward Currency**
  - Fees and rewards are primarily paid in a stablecoin (e.g., USDC, etc) however new stablecoin assets prevent value depreciation (e.g., RAI [52]) that is not pegged to centralized stablecoin assets are worth exploring.
  - This approach helps mitigate risks associated with inflationary measures or other external factors affecting centralized stablecoins, to fairly reward Participants.
  - The use of native tokens in lieu of stablecoins remains an open question for further exploration.

- **Evaluator Staking and NFTs**
  - Evaluators must stake an NFT to participate in evaluation processes i.e., to earn network fees.
  - These NFTs represent the Evaluator's stake and reputation in the network.
  - NFT values are self-assessed and based on the chosen stablecoin to ensure stability.

- **NFT Valuation and Taxation**
  - NFT values are subject to **Harberger taxation:**
    - Owners periodically self-assess their property and pay tax on its value.
    - Others are able to purchase the property from the owner at the taxed price at any time, forcing a sale [53].
  - Harberger taxation is priced in the fees paid to Evaluators (i.e., stablecoin, native token,etc) and is charged periodically based on the value of the owner's NFT asset.
  - Values can be adjusted either:
  a) Dynamically based on performance, or
  b) At set periods, after which they become open to auction (PCO mechanism).

- **Evaluator Incentive Structure**
  - As Evaluators perform more work, they:
  a) Receive more rewards from fees.
  b) Can assess their NFT stake at a higher value, as the profits from fees exceed the amount taxed.
  - This structure incentivizes high-quality evaluations and active participation [53].

- **Partial Common Ownership (PCO)**
  - Implements a mechanism where NFTs can be put up for auction after certain periods or at all times.
  - Helps maintain fair valuation and prevents monopolistic behavior; setting non-speculative asset pricing which reflects work done in the network.

**This protocol aims to create a balanced, fair, and efficient system for decentralized machine learning model evaluation and improvement. It is also a novel proposal for the implementation of a public goods funding mechanism.** By leveraging economic incentives and novel ownership structures, the protocol aligns the interests of all participants towards the common goal of advancing AI capabilities, while simultaneously contributing to the funding of public goods. It addresses potential challenges and areas for further refinement, ensuring a sustainable and equitable ecosystem for all involved. Alternatively, it considers leveraging existing ERC standards or exploring other traditional staking protocols—such as those used with a native coins—that do not aim to become part of a public goods ecosystem to meet the wider protocol requirements [49].

**Public Goods & Governance in *DIN*:** Projects like Gitcoin funding vision for community-driven proposals and public goods funding highlight the increasing focus on supporting shared resources and communal benefits [54], [55]. The proceeds from the taxation mechanism within our protocol are **allocated towards funding *DINs* open-source public goods infrastructure like or broader ecosystems**. This could involve supporting the network's own public goods or contributing to initiatives such as Gitcoin grants. **These approaches align with principles embraced by communities such as RadicalxChange (RxC), the Plurality Book, Ethereum blockchain ecosystems, and the Kernel Community. They also reflect the values of RDI Berkeley in DeAI, which emphasize openness, responsibility, and a democratized AI economy.** By integrating transparent, community-driven mechanisms and decentralized models, these approaches promote equitable participation and resource distribution [53], [56], [57], [58], [59]. Their work underscores the need for **ongoing experimentation and a willingness to explore new ideas**, crucial for developing transparent systems that benefit the public.

*DIN* is an organizational network and public goods software by design, though its applications extend beyond the realm of public goods. It emphasizes self-sovereignty and may explore governance mechanisms to keep engaged, incorporating models such as Gov4Git (non-coin-based voting) and quadratic voting tools to engage contributors [60]. The concept of sovereign networks is particularly relevant to personal data stores, as it delves into decentralized governance, commons-based peer production, and digital communities with shared values that operate independently of traditional structures. These sovereign networks explore the potential for decentralized models to reshape governance and resource distribution in novel ways, and may also contain relevance to experimental network states [61], [62].

*DIN* may include **DPPs** (*DIN* Proposal Protocols) for drawing attention to proposals for improving the network. These proposals will be voted upon by contributors who are allotted non-coin voting credits, employing tools akin to those or including Gov4Git [60]. A flexible, inclusive rollout driven by community input is essential to mitigate wealth concentration within the crypto ecosystems. Circulating financial value within ecosystems that benefit the public can stimulate economically advantageous societies. This approach, coupled with a commitment to continuous innovation, is vital for advancing these concepts in dynamic and impactful ways.

## 9. Conclusion & Future Works

*Decentralized Intelligence Network (DIN)* is a theoretical framework designed to address challenges in AI development and deployment, particularly focusing on data fragmentation and siloing issues. Its core objective is to enable scalable AI through data sovereignty while facilitating effective AI utilization within sovereign networks. DIN represents a significant advancement in integrating key themes of:

- Data sovereignty
- Public blockchain
- Decentralized federated learning (FL)
- Off-chain file storage (e.g., IPFS)
- Reward protocols

By introducing a decentralized FL protocol within a sovereign architecture, DIN enables individuals to retain ownership and control over their data while receiving rewards for its use. This scalable framework addresses the limitations of siloed data, benefiting both participants and data users, and includes a robust, decentralized auditing system for equitable reward distribution, without third party - maintaining our requirements of sovereignty.

**Potential Benefits:**

- **Enhanced AI Performance:** Access to more diverse data can lead to more robust and capable AI systems.
- **Improved Data Utilization:** Allows for more effective use of existing data sources that may currently be underutilized due to fragmentation or siloing.
- **Preserved Privacy and Control:** Maintains data sovereignty while still enabling collaborative AI development.
- **Scalable Solutions:** Provides a framework for AI systems to grow and adapt as more data sources become available.
- **Future of an open, responsible AI economy**: AI agents automate tasks, giving humans more time to be human. The value produced by the AI economy lifts the standard of living for everyone, with safe applications of AI enabling scientific discovery and real-world sector applications, leading to a post-scarcity society.

*DIN* can potentially change how AI is developed and deployed, especially in scenarios where data privacy, sovereignty, and diverse data access are crucial factors. It's important to note that *DIN* is a systems architecture for a network and does not address other subsections of Decentralized AI (DecAI). Institutional leverage of data whilst preserving interests (move beyond private setups, preventing need for collaboration, enabling wider monetization strategies whilst preserving privacy). New economic models for tokenized systems with circular economies, with specific use case applications cross-sector inc. but not limited to finance, smart cities, healthtech, education, and more. Both decentralized and traditional institutions may benefit from this technology.

**Future Works:** While there are challenges associated with implementation, technological advancements provide a strong foundation for ongoing development. Future enhancements to DIN may involve:

- Expanding interactions with sovereign data stores
- Integrating more advanced, computationally efficient privacy-preserving techniques
- Addressing implementation complexities
- Optimizing performance trade-offs
- Improving scalability issues

These advancements will be crucial in continuing to evolve decentralized FL frameworks effectively.

**Call To Action:** *DIN* encourages researchers, practitioners, and stakeholders to engage with this framework to promote data ownership and individual sovereignty. Collaborative efforts can lead to scalable, sovereign data solutions that advance technology while respecting individual data rights.

By working together, we can overcome the challenges associated with decentralized intelligence networks and create a future where AI development is both powerful and respectful of individual privacy and data ownership.

## 11. Acknowledgments

Special thanks to Paritosh Ramanan, Rui Zhao, Harry Cai, Peng "Dana" Zhang, and Jesse Wright for their invaluable discussions on many of these ideas. Specifically, their contributions include scalable FL architectures (Paritosh), scalable decentralized auditing protocol (Paritosh, Harry), sovereign architectures, and decentralized identity management (Dana). Special thanks to Rui and Jesse, who provided significant insights across various aspects of the framework. Their insights and contributions have significantly shaped the development of this framework. Additionally, these works have been selected for presentation as a speaker at the **Summit on Responsible Decentralized Intelligence - Future of Decentralization and AI**, hosted by **Berkeley RDI** on **August 6, 2024**, at the **Verizon Center, Cornell Tech Campus, Roosevelt Island, NYC**. This summit offers an exciting opportunity to share and further refine these ideas with a broader audience.